\documentclass[aps,showpacs,onecolumn,superscriptaddress]{revtex4}
\usepackage{graphics,color}
\usepackage{epsfig}
\usepackage{dcolumn}
\usepackage{bm}
\usepackage{epsfig}
\usepackage{mathrsfs}
\usepackage{amsfonts}
\usepackage{graphicx}
\usepackage{amsmath}
\usepackage{amssymb}

\begin{document}

\title{The representations of Temperley-Lieb algebras  and entanglement in a Yang-Baxter system}

\author{Chunfang Sun}
\address{School of Physics, Northeast Normal University,
Changchun 130024, People's Republic of China}
\author{Gangcheng Wang}
\address{School of Physics, Northeast Normal University,
Changchun 130024, People's Republic of China}
\author{Hu Tao-Tao}
\address{School of Physics, Northeast Normal University,
Changchun 130024, People's Republic of China}
\author{Chengcheng Zhou}
\address{School of Physics, Northeast Normal University,
Changchun 130024, People's Republic of China}
\author{Qing-Yong Wang}
\address{School of Physics, Northeast Normal University,
Changchun 130024, People's Republic of China}
\author{Kang Xue}  \email {Xuekang@nenu.edu.cn}
\address{School of Physics, Northeast Normal University,
Changchun 130024, People's Republic of China}

\begin{abstract}
A method of constructing Temperley-Lieb algebras(TLA)
representations has been introduced in [Xue \emph{et.al}
arXiv:0903.3711]. Using this method, we can obtain another series of
$n^{2}\times n^{2}$ matrices $U$ which satisfy the TLA with the
single loop $d=\sqrt{n}$. Specifically, we present a $9\times9$
matrix $U$ with $d=\sqrt{3}$. Via Yang-Baxterization approach, we
obtain a unitary $ \breve{R}(\theta
,\varphi_{1},\varphi_{2})$-matrix, a solution of the Yang-Baxter
Equation. This $9\times9$ Yang-Baxter matrix is universal for
quantum computing.


\end{abstract}
\vspace{0.3cm}  \maketitle

\section{Introduction}
Quantum entanglement, which has been singled out by
Schr$\ddot{o}$dinger as "the characteristic trait of quantum
mechanics" many decades ago, is the most surprising nonclassical
property of composite quantum systems. In recent years, there has
been an ongoing effort to characterize qualitatively and
quantitatively the entanglement properties, because it implies a
nonclassical nature through which we can investigate the conceptual
foundations and interpretation of quantum mechanics, and, more
importantly, it provides a fundamental resource in realizing quantum
information and quantum computers \cite{Special}, such as quantum
teleportation\cite{ben1}, superdense coding \cite{ben2}, quantum key
distribution \cite{Ekert}, and telecoloning \cite{Murao}. Besides,
in highly correlated states in condensed-matter systems such as
fractional quantum Hall liquids\cite{X.G} and
superconductors\cite{S.O,V.O}, the entanglement serves as a unique
measure of quantum correlations between degrees of freedom.

 The Yang-Baxter equation (YBE) was originated
in solving the one-dimensional $\delta$-function interaction models
by Yang \cite{yang} and the statistical models on lattices by Baxter
\cite{baxter}, and introduced to solve many quantum integrable
models by Faddeev and Leningrad Scholars \cite{Sklyanin}. Very
recently, the YBE and braiding operators have been introduced to the
field of quantum information and quantum computation, and also
provide a novel way to study the quantum entanglement
\cite{qiybe1,kauffman1,qiybe3,zkg,zg,ckg,ckg2,cxg1,S. W.
Hu,wang1,sun,wang2}. As is known, the Temperley-Lieb algebras (TLA)
\cite{Temperley} are intimately connected with braid group. In fact,
The TLA is a quotient of the group algebra of the braid group, and
any representation of TLA constructs automatically a representation
of a corresponding braid group. A unitary solution of YBE can also
be constructed from a representation of TLA via the
Yang-Baxterization approach, so TLA has been widely used in the
construction of YBE solutions \cite{R.J. Baxter,Batchelor,Y. Q. Li}.
Recently, Ref. \cite{zhang} show that TLA is found to present a
suitable mathematical framework for describing quantum
teleportation, entangle swapping, universal quantum computation and
quantum computation flow. In a very recent work, a reducible
representation of the TLA is constructed on the tensor product of
$n$-dimensional spaces \cite{Kulish}. Then we expanded Kulish's
method in Ref.\cite{wang}, and we obtained a series of solutions of
TLA. With this method, another series of solutions of TLA can be
determined in this paper.

The paper is organized as follows: In Sec.\ref{sec2}, we recall the
method of constructing some $n^{2}\times n^{2}$ matrix solutions of
TLA with the single loop $d=\sqrt{n}$ which is shown in Ref.
\cite{wang}. In the following, using the same method, another series
of solutions are presented via changing some original conditions. In
Sec.\ref{sec3}. we present a $9\times9$ matrix $U$ which satisfies
the TLA with the single loop $d=\sqrt{3}$. Via Yang-Baxterization
approach, we can obtain a $9\times9$ unitary $ \breve{R}(\theta
,\varphi_{1},\varphi_{2})$-matrix, a solution of the YBE. Then we
investigate the entanglement. We show that the arbitrary degree of
entanglement for two qutrits entangled states can be generated via
the unitary $\breve{R}$-matrix acting on the standard basis. And it
is also shown that all pure entangled states of two 3-dimensional
quantum systems (\emph{i.e.}, two qutrits) can be generated from an
initial separable state via the universal $\breve{R}$-matrix if one
is assisted by local unitary transformations. In fact, we can prove
that this unitary Yang-Baxter matrix
$\breve{R}(\theta,\varphi_{1},\varphi_{2})$ is local equivalent to
the solution in Ref.\cite{cxg1}. We end with a summary.

\section{the representations of TLA}\label{sec2}
In order to keep the paper self-contained, we first briefly review
the theory of TLA \cite{Temperley}. It is a unital algebra generated
by $U_{i}$( $i=1,2,...,N-1$) which subject to the following
relations,
\begin{eqnarray}\label{1}
U_{i}^{2}=dU_{i}, ~~~U_{i}U_{i\pm1}U_{i}=U_{i},~~~
U_{i}U_{j}=U_{j}U_{i}, ~|i-j|>1
\end{eqnarray}
where $0\neq d\in\mathbb{C}$ is the single loop in the knot theory
which don't depend on the sites of the lattices. The notation
$U_{i}\equiv U_{i,i+1}$ is used, $U_{i,i+1}$ implies $1_{1}\otimes
\cdots \otimes 1_{i-1}\otimes U_{i,i+1}\otimes 1_{i+2} \otimes
\cdots \otimes1_{n}$, and $1_{j}$ represents the unit matrix of the
$j$-th particle. The TLA is easily understood in terms of
diagrammatics in Ref.\cite{kauffman2}.

In Ref.\cite{wang}, a method of constructing some $n^{2}\times
n^{2}$ matrices solutions of TLA with $n^{3}$ matrix elements has
been shown. Let us review it briefly. The representation is defined
by two invertible $n\times n$ matrices $A\in GL(n,\mathbb{C})$ and
$B\in GL(n,\mathbb{C})$, which can also be seen as an $n^{2}$
dimensional vector $A_{ab}\in (\mathbb{C}^{n}\otimes\mathbb{C}^{n})$
and an $n^{2}$ dimensional vector $B_{ab}\in
(\mathbb{C}^{n}\otimes\mathbb{C}^{n})$, respectively. The generators
$U_{i}$ can be expressed as
\begin{equation}\label{2}
(U_{i})^{ab}_{cd}=A^{a}_{b}B^{c}_{d}\in
Mat(\mathbb{C}_{i}^{n}\otimes\mathbb{C}_{i+1}^{n}),
\end{equation}
where one explicitly writes the indices corresponding to the factors
in the tensor product space $\mathcal
{H}=\otimes_{1}^{N}\mathbb{C}^{n}$ for any $n=2,3,\ldots$. The
notation $U^{ab}_{cd}\equiv U_{ab,cd}$, $A^{a}_{b}\equiv A_{ab}$ and
$B^{c}_{d}\equiv B_{cd}$ are used. In order to satisfy the second
relation of (\ref{1}), if and only if
\begin{equation}\label{3}
(BA)^{T}(AB)=I_{n\times n},
\end{equation}
and the first relation of (\ref{1}) determines the single loop $d$:
\begin{equation}\label{12}
U_{i}^{2}=U_{i}tr(A^{T}B),~~~~~~tr(A^{T}B)=d,
\end{equation}
where $tr(A^{T}B)$ denotes the trace of matrix $A^{T}B$, and $A^{T}$
denotes the transpose of matrix $A$. By means this method Eq.
(\ref{2}), one can construct lots of $n^{2}\times n^{2}$ matrix $U$
with $d=tr(A^{T}B)$ as long as $n\times n$ matrices $A$ and $B$
satisfy Eq. (\ref{3}). Especially, when $B=A^{-1}$, the generators
$U_{i}$ can be expressed as
$(U_{i})^{ab}_{cd}=A^{a}_{b}(A^{-1})^{c}_{d}$, which has been
presented in Ref. \cite{Kulish}.

For the sake of constructing some useful TLA matrices,
Ref.\cite{wang} select the nonzero elements' locations of $n\times
n$ matrix $A$ are symmetric and are the same as the $n\times n$
matrix $B$'s, and every row and every column of them have only one
nonzero element. In addition, the nonzero elements of matrices A and
B satisfy the relation
\begin{equation}\label{con2}
    B^{a}_{b}=(A^{a}_{b})^{-1} ~~~~ namely ~~~~A^{T}B=I_{n\times n}.
\end{equation}
Under this case, it is easy to see the constraint (\ref{3}) is
automatically satisfied. And one can easily verify that the single
loop $d=Tr(I_{n\times n})=n$. Then we select $n$ matrices
$U^{(i)}$($i=1,2,\ldots,n$) which can be expressed as
$(U^{(i)})^{ab}_{cd}=(A^{(i)})^{a}_{b}(B^{(i)})^{c}_{d}$, where
matrices $A^{(i)}$ and $B^{(i)}$ all satisfy above conditions. And
all their nonzero elements for $n$ matrices $A^{(i)}$ occupy
different locations. Namely, the non-vanishing matrix elements of
$A^{(i)}$ are $(A^{(i)})^{0}_{i-1}$, $(A^{(i)})^{1}_{i-2}$,
$(A^{(i)})^{2}_{i-3}$, $\cdots$ , $(A^{(i)})^{i-1}_{0}$,
$(A^{(i)})^{i}_{n-1}$, $\cdots$ , $(A^{(i)})^{n-1}_{i}$. Taking the
suns of these n matrices $U^{(i)}$ which are $n$ different solutions
of TLA, we can construct a $n^{2}\times n^{2}$ matrix solution of
TLA with $n^{3}$ matrix elements. The combined matrix $U$ reads,
\begin{equation}\label{comb}
    U=\frac{1}{\sqrt{n}}\sum_{i=1}^{n}U^{(i)}.
\end{equation}
 The limited condition which makes $U$ be a solution of TLA in Ref.\cite{wang} reads,
\begin{eqnarray}\label{limited1}
  \sum_{j=1}^{n}(B^{(i)}A^{(j)})^{T}(A^{(k)}B^{(j)}) &=& \textbf{0}_{n\times n}\nonumber
  \\,
  \sum_{j=1}^{n}(A^{(j)}B^{(i)})(B^{(j)}A^{(k)})^{T} &=& \textbf{0}_{n\times
  n},
\end{eqnarray}
where $i\neq k$ and $i,k=1,2,\cdots n$. The nonzero matrix elements
of matrices $A^{(i)}$ and $B^{(i)}$ are determined by the limited
condition Eq. (\ref{limited1}) together with their special matrix
structures. And the first one relation of (\ref{1}) determines the
single loop $d=\sqrt{n}$.

Using the same method, we can obtain another series of TLA matrices.
In this paper we select n matrices $U^{(i)}$($i=1,2,\ldots,n$),
which can be expressed as
$(U^{(i)})^{ab}_{cd}=(A^{(i)})^{a}_{b}(B^{(i)})^{c}_{d}$, where
matrix $A^{(i)}$ and matrix $B^{(i)}$ all have special matrix
structures. The nonzero elements' locations of $n\times n$ matrix
$A^{(i)}$ are antisymmetric and are the same as the $n\times n$
matrix $B^{(i)}$'s, and every row and every column of them have only
one nonzero element. In addition, the nonzero elements of matrices
$A^{(i)}$ and $B^{(i)}$ satisfy the relation
$B^{a}_{b}=(A^{a}_{b})^{-1}$, which makes the constraint (\ref{3})
be automatically satisfied. And all their nonzero elements for $n$
matrices $A^{(i)}$ occupy different locations. Namely, the nonzero
matrix elements of $A^{(i)}$ are $(A^{(i)})^{0}_{i-1}$,
$(A^{(i)})^{1}_{i}$, $(A^{(i)})^{2}_{i+1}$, $\cdots$ ,
$(A^{(i)})^{n-i}_{n-1}$, $(A^{(i)})^{n-i+1}_{0}$,
$(A^{(i)})^{n-i+2}_{1}$, $\cdots$ , $(A^{(i)})^{n-1}_{i-2}$. For
example, if n=4 and i=3, the nonzero matrix elements of $A^{(3)}$
are $(A^{(3)})^{0}_{2}$, $(A^{(3)})^{1}_{3}$, $(A^{(3)})^{2}_{0}$,
$(A^{(3)})^{3}_{1}$. Under these conditions, these n matrices
$U^{(i)}$ are $n$ different solutions of TLA with the single loop
$d^{(i)}=n$, and all their nonzero matrix elements occupy different
locations. By means of the same method as Eq. (\ref{comb}), another
combined matrix $U$ reads,
\begin{equation}\label{comb1}
    U=\frac{1}{\sqrt{n}}\sum_{i=1}^{n}U^{(i)}.
\end{equation}
We substitute Eq(\ref{comb1}) into Eqs(\ref{1}), the second relation
of Eqs(\ref{1}) determines $A^{(i)}$ and $B^{(i)}$ subject to the
same constraints as Eqs. (\ref{limited1}), so the nonzero matrix
elements of matrices $A^{(i)}$ and $B^{(i)}$ are determined by the
limited condition Eq. (\ref{limited1}) together with above special
matrix structures. And the first one relation of (\ref{1})
determines the single loop $d=\sqrt{n}$ the same as
Ref.\cite{wang}'s. So with the same method as (\ref{comb}), via
selecting the nonzero elements' locations of $n\times n$ matrices
$A^{(i)}$ and $B^{(i)}$ are antisymmetric in this paper, we can
construct another series of $n^{2}\times n^{2}$ matrices $U$ (with
$n^{3}$ matrix elements) which satisfy the TLA with the single loop
$d=\sqrt{n}$.

\section{a $9\times9$ $U$ matrix, Unitary $\breve{R}$ matrix and entanglement}\label{sec3}

In this section, we first construct a $3^{2}\times 3^{2}$ matrix $U$
which satisfies the TLA for $n=3$. Via the above summation method
$U=\frac{1}{\sqrt{3}}\sum_{i=1}^{3}U^{(i)}=\frac{1}{\sqrt{3}}\sum_{i=1}^{3}(A^{(i)})^{a}_{b}(B^{(i)})^{c}_{d}$,
which satisfies the constraints (\ref{limited1}) with above special
matrix structures, one can have the solution with standard basis
(\emph{i.e.} $|0\rangle,|1\rangle,|2\rangle$) as follows,

\begin{eqnarray}
A^{(1)}=\left(
  \begin{array}{ccc}
   1 & 0 & 0  \\
    0 &\frac{q_{1}^{2}}{q_{2}^{2}\omega} & 0 \\
    0 &0 & \frac{\omega}{q_{2}^{2}} \\
  \end{array}
  \right), ~~~~~~B^{(1)}=\left(
  \begin{array}{ccc}
    1 & 0 & 0 \\
    0 & \frac{q_{2}^{2}\omega}{q_{1}^{2}}& 0 \\
    0 & 0& \frac{q_{2}^{2}}{\omega} \\
  \end{array}
  \right)  ,~~~~~~A^{(2)}=\left(
  \begin{array}{ccc}
    0 & 1 & 0 \\
   0 & 0 & \frac{\omega}{q_{2}} \\
  \frac{1}{q_{1}\omega} & 0 & 0\\
  \end{array}
  \right) ,\nonumber
\end{eqnarray}
\begin{eqnarray}
 B^{(2)}=\left(
  \begin{array}{ccc}
    0 & 1 & 0 \\
   0 & 0 & \frac{q_{2}}{\omega} \\
  q_{1}\omega & 0 & 0\\
  \end{array}
  \right) ,~~~~~~A^{(3)}=\left(
  \begin{array}{ccc}
    0 & 0 & 1 \\
   q_{1} & 0 & 0 \\
  0 & \frac{q_{1}}{q_{2}} & 0\\
  \end{array}
  \right),~~~~~~B^{(3)}=\left(
  \begin{array}{ccc}
    0 & 0 & 1 \\
   \frac{1}{q_{1}}& 0 &0 \\
  0 & \frac{q_{2}}{q_{1}} & 0\\
  \end{array}
  \right).
\end{eqnarray}
 In this work, we choose basis $\{|00\rangle, |01\rangle,
|02\rangle, |10\rangle, |11\rangle, |12\rangle, |20\rangle,
|21\rangle, |22\rangle\}$ as the standard basis. As a result, the
$9\times 9$ Hermitian matrix $U$ with $d=\sqrt{3}$ is realized as,
\begin{eqnarray}\label{5}
U=\frac{1}{\sqrt{3}}\left(
  \begin{array}{ccccccccc}
   1 & 0 & 0 & 0 & \frac{q_{2}^{2}\omega}{q_{1}^{2}} & 0 & 0 & 0 & \frac{q_{2}^{2}}{\omega} \\
0 & 1 & 0 & 0 & 0 & \frac{q_{2}}{\omega} & q_{1}\omega & 0 & 0 \\
0 & 0 & 1 & \frac{1}{q_{1}} & 0 & 0 & 0 & \frac{q_{2}}{q_{1}} & 0 \\
0 & 0 & q_{1} & 1 & 0 & 0 & 0 &  q_{2} & 0 \\
\frac{q_{1}^{2}}{q_{2}^{2}\omega} & 0 & 0 & 0 & 1 & 0 & 0 & 0 & q_{1}^{2}\omega \\
0 & \frac{\omega}{q_{2}} & 0 & 0 & 0 & 1 & \frac{q_{1}}{q_{2}\omega} & 0 & 0 \\
0 & \frac{1}{q_{1}\omega} & 0 & 0 & 0 & \frac{q_{2}\omega}{q_{1}} & 1 & 0 & 0 \\
0 & 0 & \frac{q_{1}}{q_{2}} & \frac{1}{q_{2}} & 0 & 0 & 0 & 1 & 0 \\
\frac{\omega}{q_{2}^{2}} & 0 & 0 & 0 & \frac{1}{q_{1}^{2}\omega} & 0 & 0 & 0 & 1 \\
\end{array}
\right),
\end{eqnarray}
where $\omega=e^{\epsilon i\frac{2\pi}{3}}$(here and after
$\epsilon=\pm$), $q_{1}=e^{i\varphi_{1}}$, and
$q_{2}=e^{i\varphi_{2}}$, with the parameters $ \varphi_{1}$ and
$\varphi_{2}$ both are real. The matrix $U$ of (\ref{5}) can also be
rewrited as a form of projectors
\begin{eqnarray}
U=\sqrt{3}\sum_{i=1}^{3}|\Psi_{i}\rangle\langle\Psi_{i}|,
\end{eqnarray}
where
\begin{eqnarray}
|\Psi_{1}\rangle &=&
\frac{1}{\sqrt{3}}(|00\rangle+\frac{q_{1}^{2}}{q_{2}^{2}\omega}|11\rangle+\frac{\omega}{q_{2}^{2}}|22\rangle),\nonumber\\
|\Psi_{2}\rangle &=&
\frac{1}{\sqrt{3}}(|01\rangle+\frac{\omega}{q_{2}}|12\rangle+\frac{1}{q_{1}\omega}|20\rangle),\nonumber\\
|\Psi_{3}\rangle &=&
\frac{1}{\sqrt{3}}(|02\rangle+q_{1}|10\rangle+\frac{q_{1}}{q_{2}}|21\rangle).
\end{eqnarray}
It is interesting that all $|\Psi_{i}\rangle$($i=1,2,3$) are of the
$SU(3)$ entangled states with the maximal degree of entanglement
\cite{Kaszlikowski}.

Next we derive a unitary matrix $\breve{R}$ from $U$ by the
Yang-Baxterization approach. Such a matrix $\breve{R}$ satisfies the
YBE
\begin{eqnarray}\label{6}
\check{R}_{i}(u)\check{R}_{i+1}(\frac{u+v}{1+\beta^{2}uv})\check{R}_{i}(v)=\check{R}_{i+1}(v)\check{R}_{i}(\frac{u+v}{1+\beta^{2}uv})\check{R}_{i+1}(u),
\end{eqnarray}
where $u$ and $v$ are spectral parameters, and $\beta^{-1}=ic$ ($c$
is light velocity). The notation $\breve{R}_{i}(u)\equiv
\breve{R}_{i,i+1}(u)$ is used, $\breve{R}_{i,i+1}(u)$ implies
$1_{1}\otimes1_{2}\otimes1_{3}\cdots \otimes
\breve{R}_{i,i+1}(u)\otimes\cdots \otimes1_{n}$, and $1_{j}$
represents the unit matrix of the $j$-th particle. The physical
meaning of $\check{R}(u)$ is two-particle scattering matrix
depending on the relative rapidity $\tanh^{-1}(\beta u)$. Let the
unitary Yang-Baxter $\breve{R}$-matrix for two qutrits be the form
\begin{equation}
\breve{R}_{i}(u)=\rho(u)[\mathbf{1}_{i}+G(u)U_{i}],
\end{equation}
where $\rho(u)$ is a normalization factor, and we can choose
appropriate $\rho(u)$ to ensure $\breve{R}(u)$ is unitary. $G(u)$ is
determined by the associated YBE (\ref{6}), and we easily get
\begin{equation}\label{11}
G(u)+G(v)+\sqrt{3}G(u)G(v)=[1-G(u)G(v)]G(\frac{u+v}{1+\beta^{2}uv}).
\end{equation}
Equation (\ref{11}) has the solution $G(u)=\frac{4i\epsilon\beta
u}{1+\beta^{2}u^{2}-2\sqrt{3}i\epsilon\beta u}$. We further
introduce the transformation
$\frac{1+\beta^{2}u^{2}+2\sqrt{3}i\epsilon\beta
u}{1+\beta^{2}u^{2}-2\sqrt{3}i\epsilon\beta u}\equiv e^{-2i\theta}$,
$\rho(u)\equiv e^{i\theta}$, where $\theta$ is real. One can easily
verify $G(u)=\frac{e^{-2i\theta}-1}{\sqrt{3}}$. Then we can obtain
the following form of the unitary Yang-Baxter matrix for two qutrits
as,
\begin{eqnarray}\label{25}
\breve{R}(\theta,\varphi_{1},\varphi_{2})=\frac{1}{3}\left(
  \begin{array}{ccccccccc}
   b & 0 & 0 & 0 & \frac{aq_{2}^{2}\omega}{q_{1}^{2}} & 0 & 0 & 0 & \frac{aq_{2}^{2}}{\omega} \\
0 & b & 0 & 0 & 0 & \frac{aq_{2}}{\omega} & aq_{1}\omega& 0 & 0 \\
0 & 0 & b& \frac{a}{q_{1}} & 0 & 0 & 0 & \frac{aq_{2}}{q_{1}} & 0 \\
0 & 0 & aq_{1} & b & 0 & 0 & 0 &  aq_{2} & 0 \\
\frac{aq_{1}^{2}}{q_{2}^{2}\omega} & 0 & 0 & 0 & b & 0 & 0 & 0 & aq_{1}^{2}\omega \\
0 & \frac{a\omega}{q_{2}} & 0 & 0 & 0 & b & \frac{aq_{1}}{q_{2}\omega} & 0 & 0 \\
0 & \frac{a}{q_{1}\omega} & 0 & 0 & 0 & \frac{aq_{2}\omega}{q_{1}} & b & 0 & 0 \\
0 & 0 & \frac{aq_{1}}{q_{2}} & \frac{a}{q_{2}} & 0 & 0 & 0 & b & 0 \\
\frac{a\omega}{q_{2}^{2}} & 0 & 0 & 0 & \frac{a}{q_{1}^{2}\omega} & 0 & 0 & 0 & b \\
\end{array}
\right),
\end{eqnarray}
where $a=-2i\sin\theta$ and $b=2e^{i\theta}+e^{-i\theta}$.

The Gell-mann matrices, a basis for the Lie algebra $SU(3)$ \cite{Pfeifer}, $\lambda_{u}$ satisfy $%
[I_{\lambda},I_{\mu}]=if_{\lambda\mu\nu}I_{\nu}
(\lambda,\mu,\nu=1,\cdot \cdot \cdot ,8)$, where
$I_{\mu}=\frac{1}{2}\lambda_{\mu}$. To the later
convenience, we denote $I_{\lambda}$ by, $I_{\pm}=I_{1}\pm iI_{2}$, $%
V_{\pm}=V_{4}\mp iV_{5}$,$U_{\pm}=I_{6}\pm iI_{7}$,
$Y=\frac{2}{\sqrt{3}}I_{8} $. In this work, we get rise to three
sets of $SU(3)$ realizations as:
\begin{eqnarray}\label{16}
(i):\left\{
\begin{array}{lll}
I_{\pm}^{(1)}=I_{1}^{\pm}I_{2}^{\pm},~~~U_{\pm}^{(1)}=U_{1}^{\pm}U_{2}^{
\pm},~~~V_{\pm}^{(1)}=V_{1}^{\pm}V_{2}^{\pm},\\
&\\
I_{3}^{(1)}=\frac{1}{3}(I_{1}^{3}+I_{2}^{3})+\frac{1}{2}%
(I_{1}^{3}Y_{2}+Y_{1}I_{2}^{3}),\\
&\\
Y^{(1)}=\frac{1}{3}(Y_{1}+Y_{2})+\frac{2}{3}I_{1}^{3}I_{2}^{3}-\frac{1}{2}
Y_{1}Y_{2};
\end{array}
\right.
\end{eqnarray}
\begin{eqnarray}\label{17}
(ii):\left\{
\begin{array}{lll}
I_{\pm}^{(2)}=U_{1}^{\pm}V_{2}^{\pm},~~~U_{\pm}^{(2)}=V_{1}^{\pm}I_{2}^{%
\pm},~~~V_{\pm}^{(2)}=I_{1}^{\pm}U_{2}^{\pm} , \\
& \\
I_{3}^{(2)}=\frac{1}{2}[-\frac{1}{3}(I_{1}^{3}+I_{2}^{3})+\frac{1}{2}%
(Y_{1}-Y_{2})+I_{1}^{3}Y_{2}+Y_{1}I_{2}^{3}], \\
&\\
Y^{(2)}=-\frac{1}{3}(I_{1}^{3}-I_{2}^{3})-\frac{1}{6}(Y_{1}+Y_{2})+\frac{2}{3%
}I_{1}^{3}I_{2}^{3}-\frac{1}{2}Y_{1}Y_{2} ;&\\
\end{array}
\right.
\end{eqnarray}
\begin{eqnarray}\label{18}
(iii):\left\{
\begin{array}{lll}
I_{\pm}^{(3)}=V_{1}^{\pm}U_{2}^{\pm},~~~U_{\pm}^{(3)}=I_{1}^{\pm}V_{2}^{%
\pm},~~~V_{\pm}^{(3)}=U_{1}^{\pm}I_{2}^{\pm}, \\
&\\
I_{3}^{(3)}=\frac{1}{2}[-\frac{1}{3}(I_{1}^{3}+I_{2}^{3})-\frac{1}{2}%
(Y_{1}-Y_{2})+I_{1}^{3}Y_{2}+Y_{1}I_{2}^{3}],\\
&\\
Y^{(3)}=\frac{1}{3}(I_{1}^{3}-I_{2}^{3})-\frac{1}{6}(Y_{1}+Y_{2})+\frac{2}{3}%
I_{1}^{3}I_{2}^{3}-\frac{1}{2}Y_{1}Y_{2}.\\
\end{array}
\right.
\end{eqnarray}
We denote $I^{(k)}_{\pm}=I^{(k)}_{1}\pm iI^{(k)}_{2}$, $V^{(k)}_{%
\pm}=V^{(k)}_{4}\mp iV^{(k)}_{5}$,$U^{(k)}_{\pm}=I^{(k)}_{6}\pm
iI^{(k)}_{7}$, $Y^{(k)}=\frac{2}{\sqrt{3}}I^{(k)}_{8}$$(k=1,2,3)$.
These realizations satisfy
the commutation relation $[I^{(i)}_{\lambda},I^{(j)}_{\mu}]=i\delta_{ij}f_{%
\lambda\mu\nu}I^{(i)}_{\nu} (\lambda,\mu,\nu=1,\cdot \cdot \cdot
,8;i,j=1,2,3)$.

For $i$-th and $(i+1)$-th lattices, $\breve{R}$ can be expressed in
terms of the above operators,
\begin{eqnarray}
\breve{R}(\theta,\varphi_{1},\varphi_{2})&=&\frac{a}{3}[\frac{q_{2}^{2}\omega}{q_{1}^{2}}I_{+}^{(1)}+\frac{q_{1}^{2}}{q_{2}^{2}\omega}I_{-}^{(1)}+\frac{\omega}{q_{2}^{2}}V_{+}^{(1)}
+\frac{q_{2}^{2}}{\omega}V_{-}^{(1)}+q_{1}^{2}\omega U_{+}^{(1)}+\frac{1}{q_{1}^{2}\omega}U_{-}^{(1)} \nonumber \\
&+&\frac{q_{1}}{q_{2}\omega}I_{+}^{(2)}+\frac{q_{2}\omega}{q_{1}}I_{-}^{(2)}+\frac{q_{2}}{\omega}V_{+}^{(2)}%
+\frac{\omega}{q_{2}}V_{-}^{(2)}+\frac{1}{q_{1}\omega}U_{+}^{(2)}+q_{1}\omega U_{-}^{(2)}  \nonumber \\
&+&\frac{q_{1}}{q_{2}}I_{+}^{(3)}+\frac{q_{2}}{q_{1}}I_{-}^{(3)}+q_{2}V_{+}^{(3)}
+\frac{1}{q_{2}}V_{-}^{(3)}+\frac{1}{q_{1}}U_{+}^{(3)}+q_{1}U_{-}^{(3)}]+\frac{b}{3}(I\otimes
I).
\end{eqnarray}
 So we can say the whole tensor space
$\mathbb{C}^{3}\otimes\mathbb{C}^{3}$ is completely decomposed into
three subspaces. i.e. $
\mathbb{C}^{3}\otimes\mathbb{C}^{3}=\mathbb{C}^{3}\oplus
\mathbb{C}^{3}\oplus \mathbb{C}^{3}$. In addition, each block of
$\breve{R}$-matrix can be represented by fundamental representations
of SU(3) algebra.

When one acts $\breve{R}(\theta,\varphi_{1},\varphi_{2})$ on the
separable state $|mn\rangle$ , he yields the following family of
states
$|\psi\rangle_{mn}=\sum_{ij=00}^{22}\breve{R}^{ij}_{mn}|mn\rangle$(m,n=0,1,2).
For example, if m=1 and n=1,
\begin{equation}
|\psi\rangle_{11}=\frac{1}{3}(\frac{aq_{2}^{2}\omega}{q_{1}^{2}}|00\rangle+b|11\rangle+
\frac{a}{q_{1}^{2}\omega}|22\rangle)
\end{equation}
By means of concurrence, we study these entangled states. In Ref.
\cite{Albeverio}, the generalized concurrence (or the degree of
entanglement \cite{Hill}) for two qudits is given by,
\begin{equation}
   C=\sqrt{\frac{d}{d-1}(1-I_{1})},
\end{equation}
where
$I_{1}=Tr[\rho_{A}^{2}]=Tr[\rho_{B}^{2}]=|\kappa_{0}|^{4}+|\kappa_{1}|^{4}+\cdots+|\kappa_{d-1}|^{4}$,
with $\rho_{A}$ and $\rho_{B}$ are the reduced density matrices for
the subsystems, and $\kappa_{j}$'s($j=0,1,\ldots,d-1$) are the
Schmidt coefficients. Then we can obtain the generalized concurrence
of the state $|\psi\rangle_{11}$ as
\begin{equation} \label{20}
C=\frac{2\sqrt{2}|\sin\theta|}{3}\sqrt{1+2\cos^{2}\theta}.
\end{equation}
When $\theta= \frac{\pi}{3}$, the state $|\psi\rangle_{11}$ becomes
the maximally entangled state of two
qutrits as $|\psi\rangle_{11}=\frac{-i}{\sqrt{3}}%
(\frac{q_{2}^{2}\omega}{q_{1}^{2}}|00\rangle+e^{i\frac{2\pi}{3}}|11\rangle+\frac{1}{q_{1}^{2}\omega}|22\rangle)$.
In general, if one acts the unitary Yang-Baxter matrix
$\breve{R}(\theta,\varphi_{1},\varphi_{2})$ on the basis
${|00\rangle,|01\rangle,|02\rangle,|10\rangle,|11\rangle,|12\rangle,|20\rangle,|21\rangle,|22\rangle}$,
he will obtain the same concurrence as Eq.(\ref{20}). It is easy to
check that the concurrence ranges from 0 to 1 when the parameter
$\theta$ runs from 0 to $\pi$. But for $\theta \in [0,\pi]$, the
entanglement is not a monotonic function of $\theta$. And when
$\theta= \frac{\pi}{3}$, he will generate nine complete and
orthogonal maximally entangled states of two qutrits. The
entanglement doesn't dependent on the parameters $\varphi_{1}$ and
$\varphi_{2}$. So one can verify that parameter $\varphi_{1}$ and
$\varphi_{2}$ may be absorbed into a local operation.

In fact, we can introduce a local unitary transformation
$P=P_{1}\otimes P_{2}$, where
$P_{1}=diag\{\frac{q_{1}\omega}{q_{2}},\omega,q_{1}\}$ and
$P_{2}=diag\{\frac{q_{1}}{q_{2}},\omega,q_{1}\}$. By means of this
local transformation
$\breve{R}(\theta)=P\breve{R}(\theta,\varphi_{1},\varphi_{2})P^{-1}$,
the unitary $\breve{R}(\theta,\varphi_{1},\varphi_{2})$-matrix
(\ref{25}) is local equivalent to the universal  $\breve{R}(\theta)$
matrix for $n=3$ in Ref.\cite{cxg1}, where the proof of universality
for $n^{2}\times n^{2}$ Yang-Baxter matrix is presented. So the same
as the property of $\breve{R}(\theta)$ matrix in Ref.\cite{cxg1}, we
can also say all pure entangled states of two 3-dimensional quantum
systems (\emph{i.e.}, two qutrits) can be generated from an initial
separable state via the universal
$\breve{R}(\theta,\varphi_{1},\varphi_{2})$-matrix (\ref{25}) if one
is assisted by local unitary transformations.

\section{Summary}\label{sec4}
In this paper, using the method which is shown in Ref.\cite{wang},
we can obtain another series of $n^{2}\times n^{2}$ matrices $U$
which satisfy the TLA via changing some original conditions. The
single loop of these matrices $U$ is $d=\sqrt{n}$. Then we present a
$9\times9$ matrix representation $U$ which satisfies the TLA with
the single loop $d=\sqrt{3}$, and we derived a unitary
$\breve{R}(\theta,\varphi_{1},\varphi_{2})$-matrix via
Yang-Baxterization of the $U$-matrix. Finally, we investigate the
entanglement properties of $\breve{R}$-matrix, and it is shown that
the arbitrary degree of entanglement for two-qutrit entangled states
can be generated via the unitary matrix $\breve{R}$-matrix acting on
the standard basis. We also show that all pure two-qurtit entangled
states can be generated via the universal $\breve{R}$-matrix
assisted by local unitary transformations.

\hspace{1cm}

 This work was supported in part by NSF of China (Grant
 No. 10875026).

\hspace{1cm}


\baselineskip 22pt

\begin{thebibliography}{99}
\bibitem{wang} K. Xue \emph{et.al.}arXiv:0903.3711.
\bibitem{Special} Special issue on quantum information, { \it Phys. World} {\bf 11} (1998) 33-57.
\bibitem{ben1} C. H. Bennett and G. Brassard, C. Cr\'{e}peau, R. Jozsa, A Peres, and W. K. Wootters, {\it Phys. Rev. Lett.} {\bf 70}, 1895(1993).
\bibitem{ben2} C H. Bennett and S. J. Wiesner, {\it Phys. Rev. Lett.} {\bf 69},
2881(1992).
\bibitem{Ekert} A. K. Ekert, {\it Phys. Rev. Lett.} {\bf 67}, 661
(1991).
\bibitem{Murao} M. Murao {\it et al.}, {\it Phys. Rev. A} {\bf 59}, 156
(1999).
\bibitem{X.G} X. G. Wen, Phys. Lett. A 300, 175 (2002).
\bibitem{S.O} S. Oh and J.
Kim, Phys. Rev. B 71, 144523 (2005).
\bibitem{V.O} V. Vedral, New J. Phys. 6,
102 (2004).














\bibitem{yang} C. N. Yang, {\it Phys. Rev. Lett.} {\bf 19}, 1312(1967); C. N. Yang, {\it Phys. Rev.} {\bf 168},
1920(1968).
\bibitem{baxter} R. J. Baxter, Exactly Solved Models in Statistical Mechanics (Academic
Press, London, 1982); R. J. Baxter, {\it Ann. Phys.} {\bf 70},
193(1972).
\bibitem{Sklyanin} E. K. Sklyanin, Zapiski Nauchnykh Seminarov
Leningradskogo Otdeleniya Matematicheskogo. Instituta im. V. A.
Stekiova AN SSSR, {\bf 95}, pp.55-128, 1980; L.D. Faddeev,
Integrable models in 1+1 dimensional QFT, Les Hounches Lectures, pp.
536-608, Elsevier, Amsterdam, 1984; P. P. Kulish and E. K. Sklyanin,
Lecture nots in Phys. {\bf 151}, pp. 61-119.
\bibitem{qiybe1} A. Y. Kitaev, {\it Ann. Phys.} {\bf 303}, 2(2003).
\bibitem{kauffman1} L. H. Kauffman and S. J. Lomonaco Jr., {\it New J. Phys.} {\bf 6}, 134(2004).
\bibitem{qiybe3} J. M. Franko, E. C. Rowell, and Z Wang, J. Knot Theory Ramif. {\bf 15}, 413(2006).
\bibitem{zkg} Y. Zhang, L. H. Kauffman, and M. L. Ge, {\it Int. J. Quant. Inf.} {\bf 3}, 669(2005).
\bibitem{zg} Y Zhang and M. L. Ge, {\it Quant. Inf. Proc.} {\bf 6}, 363(2007);
Y. Zhang, E. C. Rowell, Y. S. Wu, Z. H. Wang, M. L. and Ge, {\it
e-print} quant-ph/0706.1761(2007).
\bibitem{ckg} J. L. Chen, K. Xue, and M. L. Ge, {\it Phys. Rev. A.} {\bf 76}, 042324(2007).
\bibitem{ckg2} J. L. Chen, K. Xue, and M. L. Ge, {\it Ann. Phys.} {\bf 323}, 2614(2008).
\bibitem{cxg1} J. L. Chen, K. Xue, and M. L. Ge, {\it e-print} quant-ph/0809.2321.
\bibitem{S. W. Hu} S. W. Hu, K. Xue, and M. L. Ge, {\it Phys. Rev. A.} {\bf 78}, 022319
(2008).
\bibitem{wang1} Gangcheng Wang, Kang Xue, Chunfeng Wu, He Liang and C H
Oh, J. Phys. A: Math. Theor. \textbf{42}(2009) 125207.
\bibitem{sun} K. Xue \emph{et.al.}arXiv:0904.0092.
\bibitem{wang2} Gangcheng Wang, Chunfang Sun, Qingyong Wang, kang Xue,
arXiv:0903.3713v1.
\bibitem{Temperley} H. N. V. Temperley and E. H. Lieb, {\it Proc. Roy.
Soc. London}, {\bf A 322}, 25 (1971).
\bibitem{R.J. Baxter} R.J. Baxter, {\it J.Stat.Phys.} {\bf 28}, 1 (1982); A. L.
Owczarek and R.J. Baxter, {\it J.Stat.Phys.} {\bf 49}, 1093 (1987);
M. T. Batchelor and M. N. Barber, {\it J. Phys. A} {\bf 23}, L15
(1990).
\bibitem{Batchelor} M. T. Batchelor and A. Kuniba, {\it J. Phys. A} {\bf 24},
2599 (1991).
\bibitem{Y. Q. Li} Y. Q. Li, {\it J. Math. Phys.} {\bf 34}, 2 (1993).
\bibitem{zhang}Yong Zhang 2006 J. Phys. A: Math. Gen. \textbf{39}
11599-11622.
\bibitem{Kulish} P. P. Kulish {\it et al.}, {\it J. Math. Phys.}
{\bf 49}, 023510 (2008).
\bibitem{kauffman2}L. H. Kauffman and S. J. Lomonaco Jr., New J. Phys.\textbf{4},73.1¨C73.18.(2002).

\bibitem{Kaszlikowski} D. Kaszlikowski, D. K. L. Oi, M.Christandl, K. Chang, A. Ekert, L. C. Kwek and C. H. Oh, {\it Phys.Rev. B} {\bf 67} (2003) 012310.
\bibitem{Pfeifer} W. Pfeifer, The lie Algebra SU(N) An Introduction,
Birkhauser Verlag (2003).
\bibitem{Albeverio} S. Albeverio and S. M. Fei, J. Opt. B: quantum
Semiclass. Opt. {\bf 3}, 223 (2001).
\bibitem{Hill} S. Hill and W. K. Wootters, {\it Phys. Rev. Lett.} {\bf 78}, 5022
(1997); W. K. Wootters, {\it Phys. Rev. Lett.} {\bf 80}, 2245
(1998).


\end{thebibliography}
\end{document}